\documentclass[preprint]{elsarticle}
\usepackage{color}
\usepackage{graphicx}
\usepackage{hyperref}
\usepackage{multirow}
\usepackage{verbatim}
\usepackage{setspace}
\usepackage{fontenc}
\usepackage{amsmath}
\usepackage{caption}
\usepackage{chemfig}
\usepackage{lineno,hyperref}
\modulolinenumbers[5]
\biboptions{numbers}
\biboptions{sort&compress}

\journal{Computational Materials Science}
\begin{document}

\begin{frontmatter}

\title{Machine Learning Models for the Lattice Thermal Conductivity Prediction of Inorganic Materials}

\author[mymainaddress]{Lihua Chen}
\author[mymainaddress]{Huan Tran}
\author[mymainaddress]{Rohit Batra}
\author[mymainaddress]{Chiho Kim}
\author[mymainaddress]{Rampi Ramprasad\corref{mycorrespondingauthor}}
\cortext[mycorrespondingauthor]{Corresponding author}
\ead{rampi.ramprasad@mse.gatech.edu}

\address[mymainaddress]{School of Materials Science and Engineering, Georgia Institute of Technology, 771 Ferst Drive NW, Atlanta, GA 30332, United States}

\begin{abstract}
	The lattice thermal conductivity ($\kappa_{\rm L} $) is a critical property of thermoelectrics,  thermal barrier coating materials and semiconductors. While accurate empirical measurements of $\kappa_{\rm L} $ are extremely challenging, it is usually approximated through computational approaches, such as semi-empirical models, Green-Kubo formalism coupled with molecular dynamics simulations, and first-principles based methods. However, these theoretical methods are not only limited in terms of their accuracy, but sometimes become computationally intractable owing to their cost. Thus, in this work, we build a machine learning (ML)-based model to accurately and instantly predict $\kappa_{\rm L}$ of inorganic materials, using a benchmark data set of experimentally measured $\kappa_{\rm L} $ of about 100 inorganic materials. We use advanced and universal feature engineering techniques along with the Gaussian process regression algorithm, and compare the performance of our ML model with past theoretical works. The trained ML model is not only helpful for rational design and screening of novel materials, but we also identify key features governing the thermal transport behavior in non-metals. 
\end{abstract}


\end{frontmatter}


\section{Introduction}

The lattice thermal conductivity ($\kappa_{\rm L}$) dictates the ability of a non-metal to conduct heat, and serves as a critical design parameter for a wide range of applications, including thermoelectrics for power generation \cite{GAYNER2016330,ZHANG201592}, thermal barrier coatings for integrated circuits \cite{cao2004ceramic,darolia2013thermal}, and semiconductors for microelectronic devices \cite{nolas2004thermal}.  Depending on the specific application, materials with different ranges of $\kappa_{\rm L}$ values are desired. For example, low $\kappa_{\rm L}$ is preferred as thermoelectrics (e.g., PbTe and Bi$_2$Te$_3$) to maximize the thermoelectric figure of merit, while for semiconductors (e.g., SiC and BP), high $\kappa_{\rm L}$ is required to avoid overheating in electronic devices. Motivated by their practical and technological significance, extensive theoretical and empirical efforts have been made to compute $\kappa_{\rm L}$, aimed at discovering materials with targeted thermal conductivity for specific applications.

In one of the early and famous theoretical works, $\kappa_{\rm L}$ of inorganic materials was estimated using semi-empirical Slack model \cite{morelli2006high}, which relies on the Debye temperature ($\Theta_{\rm D}$) and the Gruneisen parameter ($\Upsilon$) as inputs, obtained from either experimental measurements or first-principles calculations \cite{morelli2006high, toher2014high}. Although the Slack model can provide a quick $\kappa_{\rm L}$ estimate, the uncertainty in its input parameters ($\Theta_{\rm D}$, $\Upsilon$) severely impacts its prediction accuracy. Slight modifications in the functional form of the Slack model (or its closely related Debye-Callaway model \cite{DCmodel}) have also been attempted by treating certain power coefficients as fitting parameters, which are determined using experimentally measured $\kappa_{\rm L} $ values.

However, the underlying problem of $\Theta_{\rm D}$ and $\Upsilon$ uncertainty and their unavailability for new materials persists. Alternatively, the Green-Kubo formalism, combined with non-equilibrium molecular dynamics simulations, has been employed to predict $\kappa_{\rm L}$ in semiconductors (e.g., Si) \cite{PhysRevB.61.2651, PhysRevB.81.214305, Papanikolaou_2008}. However, this method can only be used for materials for which reliable atomistic force fields are available. With the recent developments of computing power and first-principles implementations, the {\it ab initio} Green-Kubo approach has been proposed to compute the $\kappa_{\rm L}$ of Si and ZrO$_{2}$, but it is limited by the high computational cost to achieve the heat flux  and system size convergences \cite{abgk}. Additionally, the phonon Boltzmann transport equation (BTE) can now be solved numerically within the relaxation time approximation \cite{seko2015prediction, broido2007intrinsic,phono3py,PhysRevLett.120.105901}. In this approach, $\kappa_{\rm L}$ is computed from the group velocity, the mode-dependent heat capacity, and the single-mode relaxation time (approximated by the phonon lifetime), all of which rely on either the harmonic or the anharmonic force constants computed at the first-principles level. While BTE calculations could in principle be done for large systems \cite{PhysRevLett.120.105901}, they are generally restricted to small unit cells owing to high computational costs.
\begin{figure}
	\centering
	\small
	\includegraphics[width=0.6\textwidth]{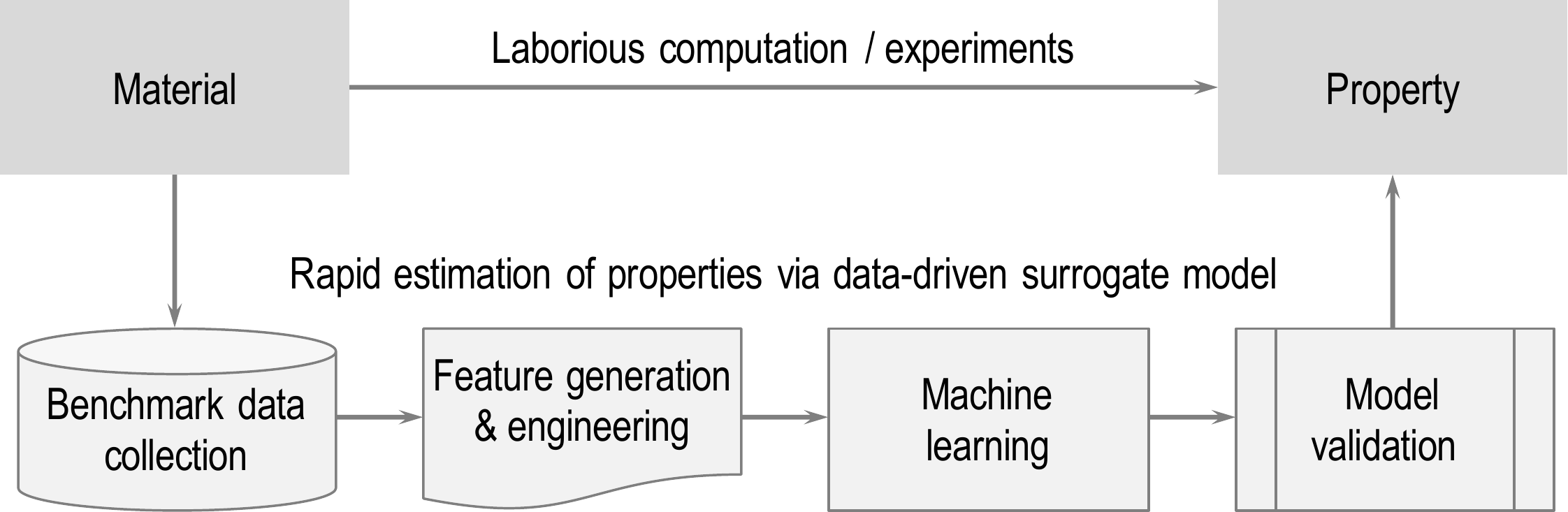}
	\caption{Schematic of the workflow adopted to build data-driven models of $ \kappa_{\rm L} $.}
	\label{fig1}
\end{figure}

Machine learning (ML) based methods, which are emerging in Materials Science and Engineering \cite{gaultois2013data,yan2015material, ramprasad2017machine, mannodi2016rational, Huan:Data, MANNODIKANAKKITHODI2017} provide yet another approach to build surrogate models to rapidly predict the thermal conductivity of materials. Seko {\it et al.} developed ML models based on $\kappa_{\rm L}$ computed for 110 materials (by solving the phonon Boltzmann transport equation as mentioned above) and a set of descriptors characterizing elemental and structural properties \cite{seko2015prediction,PhysRevB.95.144110}. The main concern with such ML model is the discrepancy between the DFT computed training data and the actual experimental values (especially for solids with very high $\kappa_{\rm L}$) which directly impacts the accuracy of these models. Furthermore, the identification of key features in determining the $\kappa_{\rm L}$ is far from trivial.

\begin{figure}[t]
	\centering 
	\includegraphics[width=0.6\textwidth]{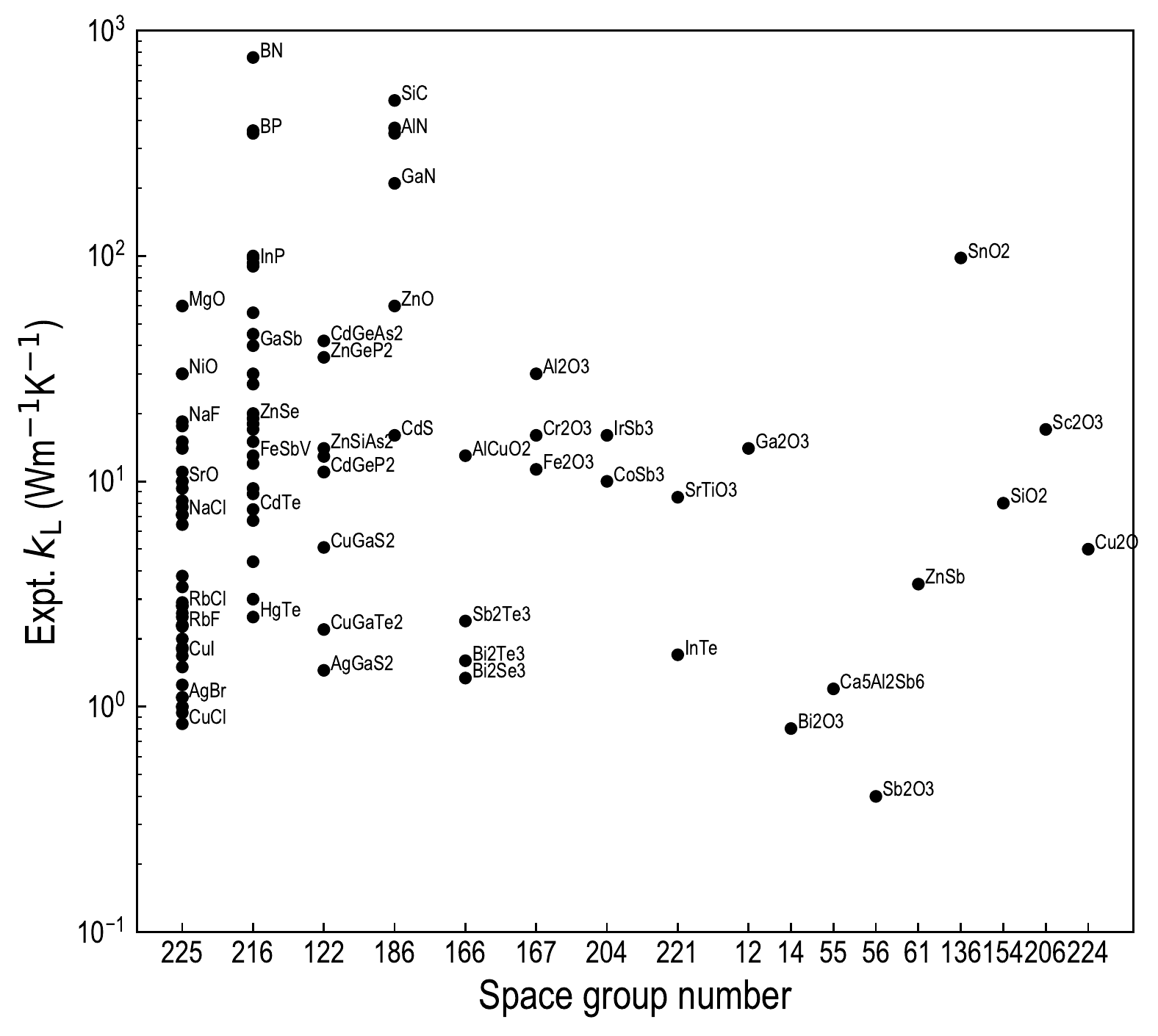}
	\caption{Experimentally measured $ \kappa_{\rm L}$ for 100 inorganic compounds with respect to their space group number. For space group 225 and 216, only a few representative cases are labeled.}
	\label{exptk}
\end{figure}
To fill the above-mentioned gaps, we have built an ML model for $\kappa_{\rm L}$, starting from a benchmark empirical data set of 100 inorganic compounds. The scheme adopted in this work is illustrated in Figure \ref{fig1}. First, the recently released Matminer package \cite{ward2018matminer} was used to generate a comprehensive list of 63 features to numerically represent the materials. This step was followed by the recursive feature elimination algorithm, down selecting the relevant features. The Gaussian process regression (GPR) algorithm, with 5-fold cross-validation (CV), was then utilized to build predictive models. The performance of the $\kappa_{\rm L}$ models was compared with past studies and validated by 5 unseen materials. The developed ML model, which is trained on the $\kappa_{\rm L}$ dataset spanning across 3 orders of magnitude, can be used to instantly predict $\kappa_{\rm L}$ of new inorganic materials while the associated GPR uncertainty could indicate whether the new materials are within the training domain or not. It is hoped that the model developed in this work can be used to screen new inorganic materials with targeted $\kappa_{\rm L}$, and it can be systematically improved when new materials are identified and added to the initial dataset.

\section{Technical Details}
\subsection{Data set}

Figure \ref{exptk} and Table \ref{dataset} summarize the dataset of empirically measured $\kappa_{\rm L}$ values (at room temperature) for 100 single crystal inorganic materials collected from the literature \cite{toher2014high, morelli2006high,yan2015material,toberer2011phonon, ref108,ref112,ref115,ref118, ref121, ref129,ref134,ref131,ref15,ref136,ref40, ref41, ref42, ref44, ref49, ref60,ref62,ref63,ref64,ref65,ref66,csbr,cscl,csi,agbr, rbf, mgse,Mg2-1,Mg2-2}, including 81 binary and 19 ternary compounds. $\kappa_{\rm L}$ of single-element materials are excluded since thermal conductivity of individual elements within a compound were used as features. The dataset is significantly diverse in chemical compositions (35 cations and 22 anions), crystal structures (with space group 225, 216, 122, 186, etc.), and the range of $\kappa_{\rm L}$, which spans over 3 orders of magnitude ($0.4 - 760 {\rm Wm}^{-1}{\rm K}^{-1} $).  The entire $ \kappa_{\rm L} $ data set---along with the bulk modulus feature values---is provided in Table S1 of the Supporting Information (SI).

Given the wide range of $\kappa_{\rm L}$, our learning problem was framed in the logarithmic scale, i.e., log($\kappa_{\rm L}$) was set as the target property, to allow better generalization of the ML models across the entire range. Furthermore, 95 out of 100 cases were used to train (with CV) the ML models, while the remaining 5 data points were held-out separately (completely unseen to the entire training process) to further validate the performance of the learned $\kappa_{\rm L}$ model. For cases where multiple $\kappa_{\rm L}$ values were reported in the literature, their average was used to train the ML model. 

\begin{table}[t]
\centering
	\small
	\caption{\small Different properties of the $\kappa_{\rm L}$ data set utilized in this work, including the class of materials, their chemical composition and space group, and the range of $\kappa_{\rm L} $ values.}
	\begin{tabular}{lcccc}
		\hline
		\hline
		Classification Area & Category                        & Examples                                    & Count \\
		\hline
		\multirow{2}{*}{Compounds}               & Binary     & AgBr, SiO$_2 $, Al$_2 $O$_3 $, ...        & 81    \\
		& Ternary  & AgGaS$_2 $,  HfCoSb, ...   & 19    \\
		\hline
		Chemical & Cations          & Na, K, Li, Be, Mg, Al, ... & 35    \\
		composition& Anions           & F, Cl, Br, I, O, S, Se, ...    & 22    \\
		\hline
		Space group        & 225              & CuCl, SnTe, NaCl, ...                                         & 39    \\
		& 216              & InSb, AlAs, SiC, ...                                          & 26    \\
		& 122, 166, etc.          &    CdGeP$_2 $, Bi$_2 $Se$_3, ... $           & 35   \\
		\hline
		Expt. $ \kappa_{\rm L} $ &0.4 -- 10&Sb$_2 $O$ _{3} $, AgCl, Mg$ _{2} $Sn, ... & 53\\
		(Wm$ ^{-1} $K$ ^{-1} $)&10 -- 100&CoO, ZnS, CdGeAs$_2 $, ... & 40\\
		at room temperature&100 -- 760& GaN, BN, BeO, ... & 7\\
		\hline
		\hline
	\end{tabular}
	\label{dataset}
\end{table}
\subsection{Feature set and dimensionality reduction}
To build accurate and reliable ML models, it is important to include relevant features that collectively capture the trends in the $\kappa_{\rm L}$ values across the different materials. The features should not only uniquely represent each material, but also be readily available to allow instant predictions for new cases. In this regard, Matminer is a good resource to easily and quickly generate features, applicable specifically to the field of materials science \cite{ward2018matminer}. In total, 61 features, belonging to three distinct categories, i.e., elemental, structural and pertaining to valence electrons, were obtained using the Matminer package \cite{ward2018matminer} by providing the chemical formula and the atomic configuration of all compounds. A total of 18 elemental properties were derived, including atomic radius, atomic mass, atom number, periodic table group and row, block, Mendeleev number, covalent radius, volume per atom from ground state, molar volume, coordination number (cn), Pauling electronegativity, first ionization energy, melting point,  boiling point, thermal conductivity, average bond length and angle of a specific site with all its nearest neighbors. Since our dataset consists of binaries and ternaries, each of these elemental feature values was obtained by taking the minimum, maximum, and weighted average over the constituting chemical species, resulting in a total of 54 elemental features. For the structural features, volume per atom, packing fraction and density were considered. These quantities were computed for the crystal structure obtained from the Materials Project database \cite{Jain2013}. Moreover, 4 features that capture the average number of valence electrons in the $s$, $ p$, $d$, and $f$ shells of the constituting elements were also included. Finally, two additional features, DFT computed bulk modulus and the space group number, were also incorporated, resulting in a 63-dimensional feature vector. The values for bulk modulus of all compounds were obtained from the Material Project database \cite{Jain2013}. As per standard ML practices, all features were scaled from 0 to 1 during model training.

To retain only the relevant features, recursive feature elimination (RFE) using linear support vector regression algorithm (with 5-fold CV) was performed on the initial 63-dimensional feature vector and the dataset of 95 training points. RFE eliminates the irrelevant features by recursively ranking feature importance and pruning the least important ones. 
In our case, it reduced the dimensionality from 63 to 29 (see Table S2 of the SI). We also used random forest algorithm for feature dimensionality reduction. In particular, we trained the data set of 95 points using 100 trees, and used the feature importance/weight to determine the relevance of the features. As discussed in Section 2 of the SI, nearly 40 features were identified to be important using the random forest method, most of which were found to be consistent to those retained from the RFE scheme discussed earlier. This provides more confidence to the RFE based dimensionality reduction step performed in this work. Overall, the 29-dimensional feature vector obtained after RFE resulted in more accurate models than the original 63-dimensional feature, as will be discussed in detail next, while a detailed comparison of the RFE and random forest methods is provided in SI.

\subsection{Gaussian Process Regression}     
The Gaussian process regression (GPR) with the radial basis function (RBF) kernel was utilized to train the ML models. In this case, the co-variance function between two materials with features $\boldsymbol{x}$ and $\boldsymbol{x'} $ is given by 
\begin{equation}\label{eq:kernel}
	k(\boldsymbol{x},\boldsymbol{x'}) =  \sigma_{f}\exp \left( -\frac{1}{2\sigma^2_{l}}||\boldsymbol{x} -\boldsymbol{x'}  ||^2 \right) + \sigma^{2}_{n}.
\end{equation}
Here, three hyper parameters $\sigma_{f}$, $\sigma_{l}$ and $\sigma_{n}$ signify the variance, the length-scale parameter and the expected noise in the data, respectively. These hyper parameters were determined during the training of the models by maximizing the log-likelihood estimate. Further, 5-fold CV was adopted to avoid overfitting. Two error metrics, namely, the root mean square error (RMSE) and the coefficient of determination ($R^2$), were used to evaluate the performance of the ML models.  To estimate the prediction errors on unseen data, learning curves were generated by varying the size of the training and the test sets. We note that the test sets were obtained by excluding the training points from the data set of 95 points. The left-out set of 5 points was completely separated from the learning process, and was used for just evaluation purposes on a few ``extrapolative'' material cases. Additionally, for each case, statistically meaningful results were obtained by averaging RMSE results over 100 runs with random training and test splits.

\section{Results and Discussion}
\begin{figure*}
	\centering
	\includegraphics[width=1.0\textwidth]{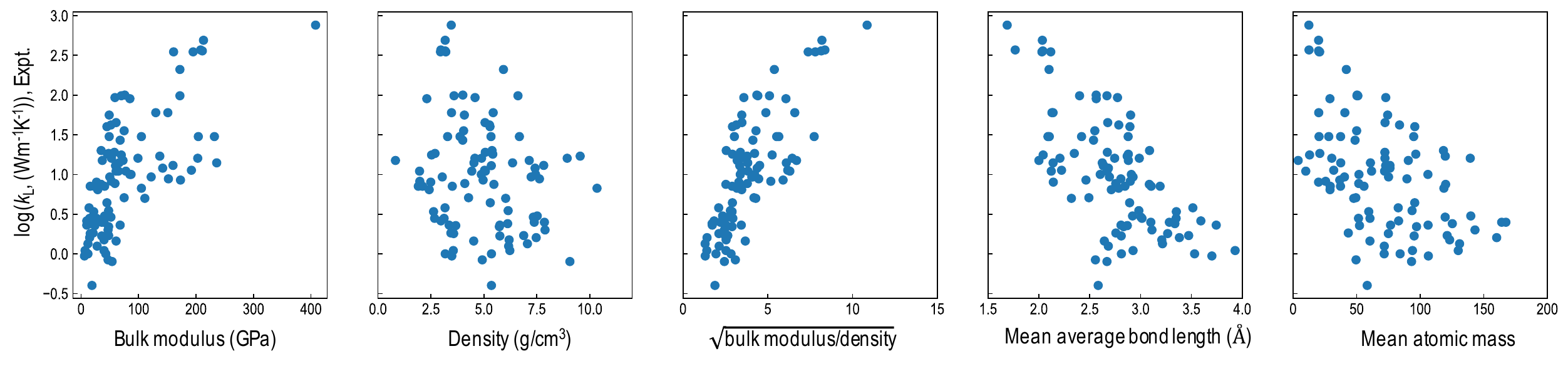}
	\caption{The correlation between experimental $ \kappa_{\rm L} $ and four representative features employed in this study. $\sqrt{\rm bulk modulus/density}$ is derived from bulk modulus and density features.}
	\label{fig5}
\end{figure*}

It is worth analyzing the correlation between these 29 features and the empirically measured $\kappa_{\rm L}$ to see how much trend is captured by these elemental, structural and chemical attributes. While in Figure \ref{fig5} we plot the $\log(\kappa_{\rm L})$ vs four important features, the corresponding plots for the remaining cases are provided in Figure S3 of SI. A strong positive correlation between $\log$($\kappa_{\rm{L}}$) and bulk modulus, and a strong inverse relation between $\log$($\kappa_{\rm{L}}$) and the mean average bond length are evident from the figure. While density alone does not show a strong correlation with $\log$($\kappa_{\rm{L}}$), the combined feature $\sqrt{\rm bulk~modulus/density}$ does indeed show a very strong linear relation. This is in-line with the physical understanding that group velocity, which is an integral part of the semi-empirical models discussed earlier, is related to the lattice anharmonic force constants, and can be approximated as $\sqrt{\rm bulk~modulus/density}$. Thus, bulk modulus can be considered to play a critical role in influencing the $\kappa_{\rm{L}}$ of different inorganic non-metals. Similarly, the inverse relationship between $\log$($\kappa_{\rm{L}}$) and the mean average bond length is also physically meaningful as when the bonds are shorter, the bond-strength anharmonicity are stronger, and the resulting $\kappa_{\rm{L}}$ is larger. For the case of mean atomic mass (a common feature used in the past ML model works), a slightly dispersed relationship is observed, indicating that it may be less important in governing $\kappa_{\rm L}$, as was the case with the rest of the 25 features illustrated in Figure S3 of the SI.

\begin{figure*}
	\centering
	\includegraphics[width=1.0\textwidth]{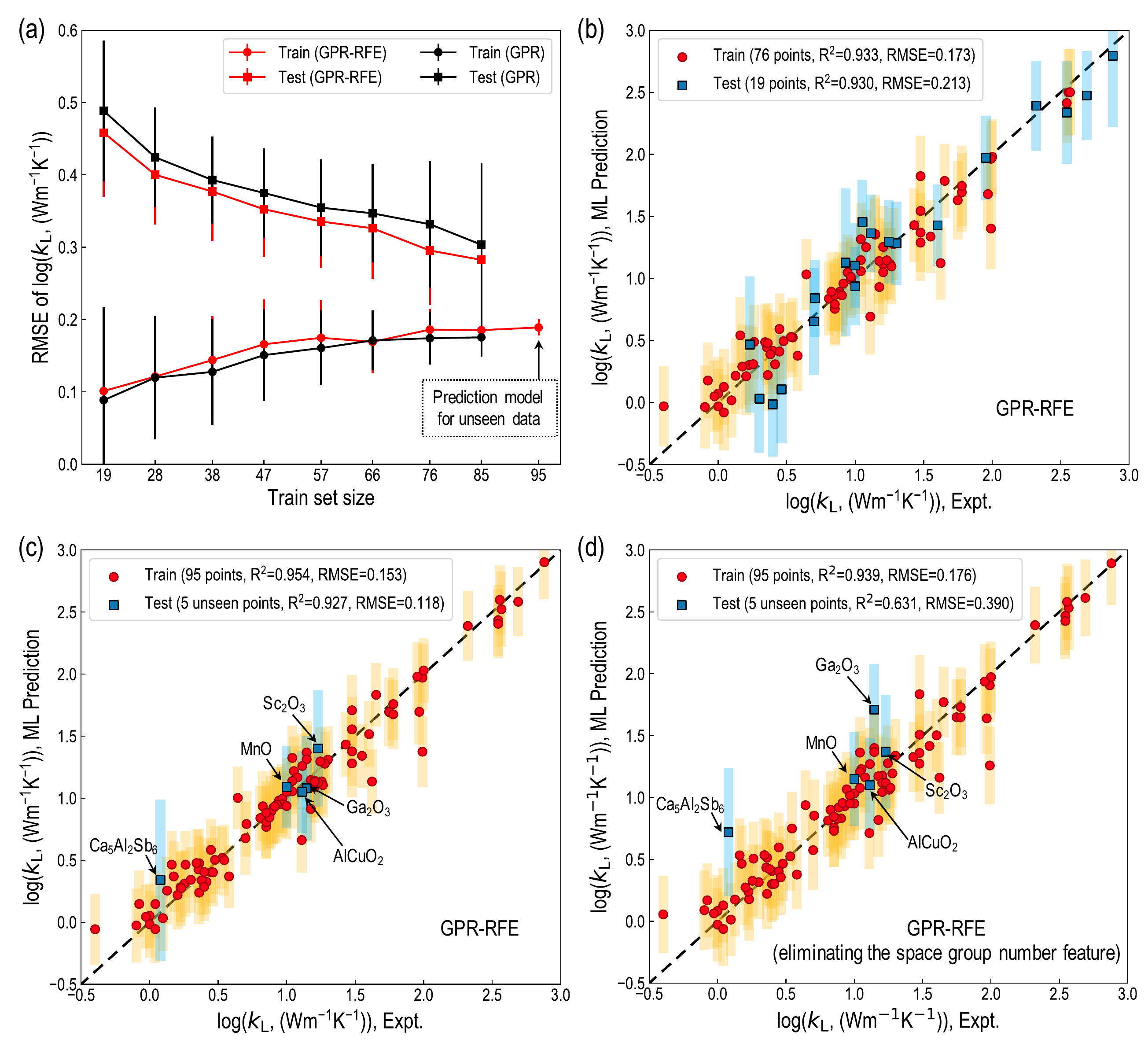}
	\caption{(a) Prediction accuracy for GPR and GPR-RFE models trained using different train set sizes, averaged over 100 runs. The corresponding test sets in (a) is the difference between total data and train sets. (b) illustrates example parity plot obtained from the GPR-RFE model (29 features) with train and test set of 76 and 19 points, respectively. Parity plots obtained from the GPR-RFE model with 95 train points and 5 unseen test points including, Sc$ _{2} $O$ _3 $, Ga$_2 $O$_3$, MnO, AlCuO$_2 $, and Ca$ _{5} $Al$ _{2} $Sb$ _{6} $, using (c) 29 features and (d) 28 features, eliminating the space group number feature from the 29 features. }
	\label{fig3}
\end{figure*}

Next, the performance of the ML models can be evaluated from the learning curves presented in Figure \ref{fig3}(a), wherein average RMSE on the training and the test sets as a function of training set size are included. The error bars denote the 1$\sigma$ deviation in the reported RMSE values over 100 runs. Results using both the initial set of 63 features (GPR), and those for the reduced 29 features (GPR-RFE) are included. Clearly, the RFE dimensionality reduction step leads to improved model performance with lower test errors, which signify better generalization of these models for unseen data. As expected, the test RMSE of both the GPR and the GPR-RFE models decreases with increase in training set size, reaching a convergence of $0.28$ in test error and of $0.18$ in train error for GPR-RFE models when the training set is about 80 \% of the data (i.e. 76 points). Figure \ref{fig3}(b) and (c) show the performance of GPR-RFE models via the example parity plots (i.e., ML predicted vs experimental log($ \kappa_{\rm L} $)), using 76 and 95 train points, respectively. The error bars in these cases represent the GPR uncertainty. Pretty high $R^{2}$ coefficient ($\geq 0.93$) on the test set in both these cases suggest a good $\kappa_{\rm L}$ model has indeed been developed.

We compared the performance of our ML model with other semi-empirical models by computing the average factor difference (AFD) \cite{DCmodel}, using the definition 
${\rm AFD}=10^a$, where $a=\dfrac{1}{N}\sum_{i=1}^{N}\left\vert{\log(\kappa_{\rm L})^{\rm expt.}-\log(\kappa_{\rm L})^{\rm model}}\right\vert$, with $N$ being the number of data points. As shown in Table \ref{afd}, the computed AFD of GPR-RFE models using the entire set of 95 points is 1.36 $\pm$0.03, which is comparable to the reported values of 1.38 and 1.46, respectively, obtained using the Slack \cite{slack} and Debye-Callaway \cite{DCmodel} models. More importantly, the latter two ML models rely on the experimental/computed features that are much more difficult to obtain owing to their dependence on the use of the Slack or Debye-Callaway models. The ML model presented here uses easily and rapidly accessible chemical and structural features derived directly from the identity of the material, making it more inexpensive and flexible. In addition, the predicted GPR uncertainty can be used to guide the next experiments via active learning \cite{kim2019active}. Further, we note that the possibility of further diversifying our ML model with data from first-principles or semi-empirical methods using multi-fidelity fusion approaches also exists \cite{mf1,mf2}.

\begin{table}
\centering
	\small
	\caption{Comparison of this work and other semi-empirical models. 1.38 and 1.46 are reported values from Slack \cite{slack} and Debye-Callaway\cite{DCmodel} models, respectively.}
	\begin{tabular}{lccc}
		\hline
		\hline
		&This work        & Slack & Debye-Callaway             \\
		\hline
		Data set         & 95          & 93                   & 55            \\
		Cross validation & 5-fold      &   leave-one-out             &4-fold\\
		Regression method&GPR-RFE &Kernel ridge&--\\
		AFD of $\kappa_{\rm L}$        & 1.36 $\pm$0.03       & 1.38                & 1.46         \\
		\hline
		\hline
	\end{tabular}
	\label{afd}
\end{table}

In order to further validate the generality and the accuracy of our ML models, we used the GPR-RFE models trained on the entire set of 95 points (see Figure \ref{fig3}(a)) to predict the log($ \kappa_{\rm L} $) of 5 unseen inorganic solids with various space group numbers present in the hold-out set. These include Sc$_2$O$_3$ (206), Ga$_2$O$_3$ (12), MnO (225), AlCuO$_2 $ (166),  and Ca$_5$Al$_2$Sb$_6$ (55), where the number within  brackets is the space group number. Figure \ref{fig3}(c) shows the comparison between the predicted and the experimental log($ \kappa_{\rm L} $), with error bars capturing the GPR uncertainty. A good performance for these 5 unseen data points is clearly evident. The high GPR uncertainty in the case of Ca$_5$Al$_2$Sb$_6$ correctly signals its space group number differences from that of the majority training data, and the application of the ML model in the ``extrapolative'' regime. Overall, the results presented here strongly advocate the good performance of the GPR-RFE models developed, which can be used to provide an inexpensive and accurate $ \kappa_{\rm L} $ prediction for other inorganic materials, especially for materials with rock-salt or zincblende structures.

Additionally, it is noteworthy that the space group number is one of the important features in our ML model, although it has little physical meaning beyond allowing the model to distinguish between different structure types. If we intentionally eliminated it from the 29 features, the test RMSE of 5 unseen materials increases from 0.12 to 0.39, as shown in Figure \ref{fig3}(d). This issue is due to the limitation of our present dataset, most of which belong to space groups 225 and 216. As a result, the space group number is required to distinguish materials in terms of their structures in the ML model. However, this problem can be solved when more data with more diverse space groups are included in the training dataset. Furthermore, our present ML model is more suitable for defect-free inorganic materials. There are some accuracy limitations of our model to predict $\kappa_{\rm L}$ of materials with defects, allotropic materials and intermetallic compounds. However, the predictive ML model can be more easily improved by actively learning on more diverse training (even temperature-dependent) data sets compared with previous semi-empirical models, due to the easily accessible features.

\section{Conclusion}

In conclusion, we have developed a simple and general ML model to predict $\kappa_{\rm L}$ of inorganic solid materials. This model is faster, and at par or more accurate than traditional physics-based computational methods. This work involves curating a benchmark dataset of experimental values of $\kappa_{\rm L}$ of 100 inorganic compounds, generating and optimizing a comprehensive set of features (using the Matminer package), and training the Gaussian Process Regression model on the data prepared. The accuracy of the developed ML models was found to be comparable to past semi-empirical models. Additionally, key features in determining $\kappa_{\rm L}$ were identified.  Overall, this present work would be useful for rational design and screening of new materials with desired $\kappa_{\rm L}$ for specific applications, and fundamentally understanding the heat transport in inorganic solid materials.
\section{Data Availability}
The entire experimental $ \kappa_{\rm L} $ data set and DFT computed bulk modulus are available in Table S1 of the Supporting Information.

\section*{Acknowledgments}
This work is supported by the Office of Naval Research through N0014-17-1-2656, a Multi-University Research Initiative (MURI) grant. 

\section*{Appendix A. Supporting Information}
\section*{Conflict of Interest:} The authors declare no competing financial interest.

\end{document}